\documentclass[11pt]{article}
\usepackage{moriond,epsfig}
\usepackage{graphicx}

\def\art{\@ifnextchar[{\eart}{\oart}}
\def\eart[#1]#2#3#4#5#6{{\rm #2}, {#3 #4} {\rm (#6) #5} [arXiv:{{#1}}]}
\def\hepart[#1]#2{{\rm #2, arXiv:{#1}}}

\newcommand{\eq}[1]{~{\rm (\ref{eq:#1})}}

\newcommand{\GeV}{\,{\rm GeV}}
\newcommand{\TeV}{\,{\rm TeV}}

\newcommand{\beq}{\begin{equation}}
\newcommand{\eeq}{\end{equation}}
\newcommand{\fig}[1]{~\ref{fig:#1}}

\def\be{\begin{equation}}
\def\ee{\end{equation}}
\def\bea{\begin{eqnarray}}
\def\eea{\end{eqnarray}}

\begin{document}
\vspace*{4cm}
\title{Implications of first LHC results}

\author{Alessandro Strumia}

\address{Dipartimento di Fisica dell'Universit{\`a} di Pisa and INFN, Italia\\
National Institute of Chemical Physics and Biophysics, Ravala 10, Tallin, Estonia}

\maketitle\abstracts{
We discuss implications of first LHC results for models motivated by the hierarchy problem: large extra dimensions and supersymmetry.
We present bounds, global fits and  implications for naturalness.}

\section{Introduction}
The main goal of the LHC is telling us why the weak scale is much below the Planck scale:
this hierarchy problem was adopted in the past 30 years as guideline of many theoretical works.
Maybe LHC will tell which CMSSM parameters are right. 
Or maybe LHC will tell which SUSY model is right. 
Or maybe LHC will tell which solution to the hierarchy problem is right. 
Or maybe LHC will tell that the hierarchy problem is not a good guideline.

The way to make progress is searching for the signals predicted by tentative solutions.
We here discuss implications of first LHC results on two of these proposals: supersymmetry and large extra dimensions.

\section{Large extra dimensions}
The hierarchy problem can be solved assuming that the quantum gravity scale $M_D$ is around the weak scale,
and that the larger Planck scale arises because gravitons live in $\delta$ extra dimensions~\cite{add}.
The following unavoidable collider signals of this scenario have been proposed:
\begin{enumerate}
\item Graviton emission (accompanied by a jet to tag the event).
\item Virtual graviton exchange, which gives the dimension 8 operator 
${\cal L}_{\rm eff} ={\cal L}_{\rm SM} + 8 {\cal T}/M_{\cal T}^4$,
$M_{\cal T}$ is expected to be comparable to $M_D$,
${\cal T} = T_{\mu\nu}^2/2$, and $T_{\mu\nu}$ is the SM energy-momentum tensor.

\item Virtual graviton exchange at one loop level, which gives the dimension 6 operator $\Upsilon = (\sum_f \bar f\gamma_\mu\gamma_5 f)^2/2$.
\item The previous signals are computable at low energy, $E\ll M_D$.  Other computable signals (such as black-hole production) arise  in the opposite limit $E\gg M_D$, which is presumably not relevant for LHC.
\end{enumerate}
In view of the high dimensionality of the operator ${\cal T}$, its effect grows  fast with energy such that LHC
(thanks to its increased energy) is more sensitive than all previous colliders, already with integrated luminosity
much lower than previous colliders.
The operator ${\cal T}$ contributes to
various cross sections:
$$ \sigma = \left(\frac{2\,{\rm TeV}}{M_{\cal T}}\right)^8 \times\left\{\begin{array}{ll}
12.5\,{\rm pb} & \hbox{for $pp\to jj$}\\
10.4\,{\rm fb}  & \hbox{for $pp\to \mu^+\mu^-$}\\
21.3 \,{\rm fb} & \hbox{for $pp\to \gamma\gamma$}\\
\end{array}\right.$$
The $pp\to jj$ channel~\cite{jj} was ignored because jets have more background than leptons or photons;
but it has a much larger cross-section, and this is the most important aspect at LHC startup,
with poor integrated luminosity.
As summarized in table~1, already with $3$/pb LHC data about $pp\to jj$
provided the dominant constraint~\cite{LHCd}.
\medskip

Since quantum corrections to the higgs mass are made finite by unknown aspects of quantum gravity, we cannot tell if the LHC bound
$M_{\cal T}>3.4\TeV$ got so strong  that large extra dimensions   no longer are a good solution to the hierarchy problem.

\begin{table}
\begin{center}
\begin{tabular}{|c|c|cc|}
\hline\hline
Experiment&Process & + & $-$\\ 
\hline
LEP & $e^+e^-\to \gamma \gamma$ & 0.93\TeV&1.01\TeV\\
LEP & $e^+e^-\to  e^+e^-$ & 1.18\TeV & 1.17\TeV\\
CDF &  $p\bar p\to e^+e^-,\gamma \gamma$& 0.99\TeV & 0.96\TeV\\ 
D\O & $p\bar p\to e^+e^-,\gamma \gamma$ & 1.28\TeV & 1.14\TeV\\
D\O &  $p\bar p\to jj$& 1.48\TeV& 1.48\TeV\\ 
CMS at 7 TeV with 34/pb & $pp\to \gamma\gamma$ &1.72\TeV&1.70\TeV\\
CMS at 7 TeV with 40/pb & $pp\to \mu^-\mu^+$ &1.6\TeV&1.6\TeV\\
ATLAS at 7 TeV with 36/pb & $pp\to jj$ & 4.2\TeV&3.2\TeV\\
ATLAS at 7 TeV with 3.1/pb &  $pp\to jj$ & 2.2\TeV&2.1\TeV\\
CMS at 7 TeV with 36/pb &  $pp\to jj$ & 4.2\TeV&3.4\TeV\\
\hline
\end{tabular}\caption{Limits on virtual graviton exchange for positive and negative interference with the SM amplitude.}
\end{center}
\end{table}

\section{Supersymmetry}
In supersymmetry quantum corrections to the higgs mass are made finite by sparticle loops.
Thereby there is neat connection between sparticle masses and the weak scale:
\begin{equation}\label{eq:MZ}
{ M^2_Z \approx 0.2 m^2_0 + 0.7 M_{3}^2-2 \mu^2}= (91\GeV)^2\times  80(\frac{M_3}{1\TeV})^2+\cdots
\end{equation}
where we assumed the CMSSM  (Constrained Minimal Supersymmetric Standard Model) and fixed $\tan\beta=3$, $A_0 = 0$, like in the experimental analyses
that presented first LHC bounds on such supersymmetric model~\cite{SUSYLHC}.

The order one coefficient of $M_3^2$ arises due to RGE running from the GUT scale down to the weak scale,
and thereby is generic of models where SUSY breaking is present at such high scale.

As a consequence of the LHC bound on the gluino mass $M_3$, its contribution to $M_Z^2$ is almost 100 times
too large, and needs to be canceled by other terms giving rise to a fine-tuning problem.
Within the CMSSM, supersymmetry allows to reduce the SM higgs mass fine tuning from $\sim 10^{30}$  down to $\sim 100$ only:
no longer down to $\sim 1$, casting doubts on the ideology according to which the hierarchy problem is solved identifying
the weak scale with  the SUSY-breaking scale.

\begin{figure}
$$\includegraphics[width=\textwidth,height=6cm]{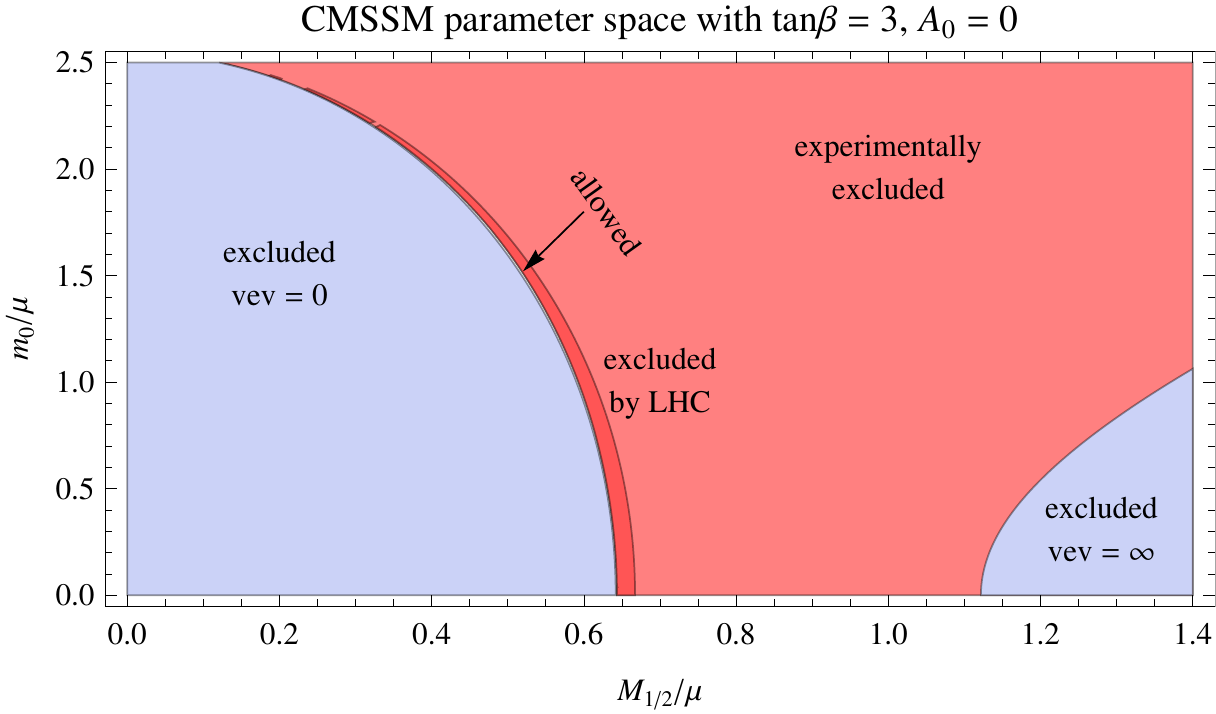}$$
\caption{An example of the parameter space of the CMSSM model. The white region is allowed.
The dashed line around the boundary of the allowed region is the prediction of the BS model.
\label{fig:RR}}
\end{figure}

To be more precise, eq.\eq{MZ} can be used to eliminate one parameter, such that the CMSSM model  has
two free  parameters.
Instead of the usual choice, $m_0$ and $ M_{1/2}\approx M_3/2.6$, 
we choose any two adimensional ratios, such that the overall scale of supersymmetry
is fixed by eq.\eq{MZ}, rather than by hand.
Only in the part of the parameter space where cancellations happen, sparticles are heavier than $M_Z$
(their natural scale according to eq.\eq{MZ}) and compatible with experimental bounds.
The result is shown in fig.\fig{RR} from~\cite{LHCnat}
(here updated at the light of the latest ATLAS data with 165/pb~\cite{SUSYLHC2}).

\begin{itemize}
\item The light-gray regions are theoretically excluded because the minimum of the potential is not the physical one.

\item The red region in the middle is theoretically allowed, but it has now been experimentally excluded.
The darker red shows the new region excluded by LHC with respect to the previous
LEP bounds.
\item
The white region is allowed, but is now so small that enlarging the picture is needed to see it.
It is close to the boundary where $M_Z=0$ and thereby has $M_Z\ll m_0,M_{1/2},\mu$.
\end{itemize}

So far we fixed $A_0=0$ and $\tan\beta=3$, but the situation within this slice of parameter space is representative
of the situation present in the full CMSSM parameter space.  To explore it we scan all its adimensional parameters,
and compute the allowed ``fraction of parameters space'',
which generalized the ``size of the allowed region'' in fig.\fig{RR}.
We find that only about $0.7\%$ of the CMSSM parameter space remains allowed.
Furthermore  non-minimal Higgs  models invented to ameliorate
the analogous issue already present after LEP no longer work,
just because the LHC bound has nothing to do with the higgs (see table~2).

LEP excluded sparticles around the $Z$ mass.
LHC now excludes heavier sparticles, and reaches the next milestone: sparticles
a loop factor above $M_Z$.
Indeed the dashed line in fig.\fig{RR} is the prediction of a model where the usual relation, eq.\eq{MZ}, gets supplemented by a relation
that demands sparticle to be a loop factor above $M_Z$: $m_{\tilde{t}}\approx 4\pi M_Z/\sqrt{12}$~\cite{BS}.

\begin{table}[b]
$$\begin{array}{c|ccc}
\hbox{experimental} & \multicolumn{3}{c}{\hbox{fraction of surviving CMSSM parameter space}}\\ 
\hbox{bound} &\hbox{any $m_h$} & m_h >100\GeV & m_h >110\GeV\\  \hline
\hbox{LEP} & 10\% & 3\% & 1\%\\
\hbox{LHC} & 2.2\% &1.2\% & 0.7\% \\
\end{array}$$
\caption{\label{tab:tab}Fraction of the CMSSM parameter space that survives to the various bounds.}
\end{table}

\bigskip

Ignoring the naturalness issue,
first data from LHC also affect CMSSM global fits~\cite{fit}, where the scale of supersymmetry is now fixed
by  i) fitting the anomaly in the $g-2$ of the muon compatibly with other indirect data; 
ii)  demanding that the thermal abundance of the lightest neutralino equals the Dark Matter abundance.

\begin{figure}
$$\includegraphics[width=0.3\textwidth]{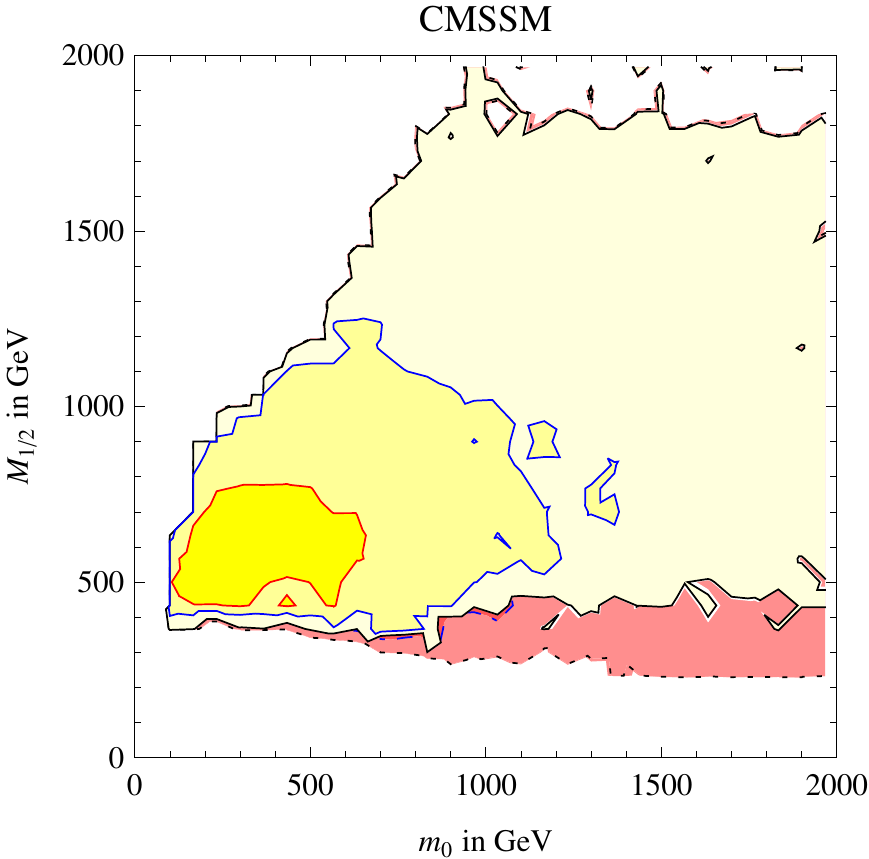}\qquad
\includegraphics[width=0.3\textwidth]{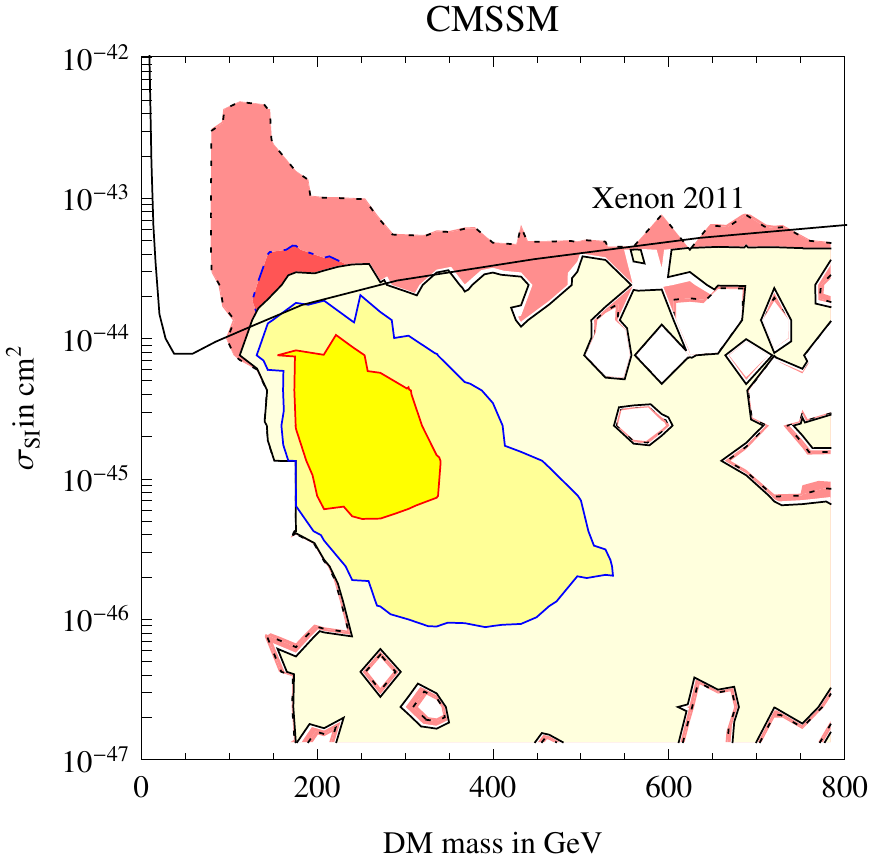}\quad
\includegraphics[width=0.3\textwidth]{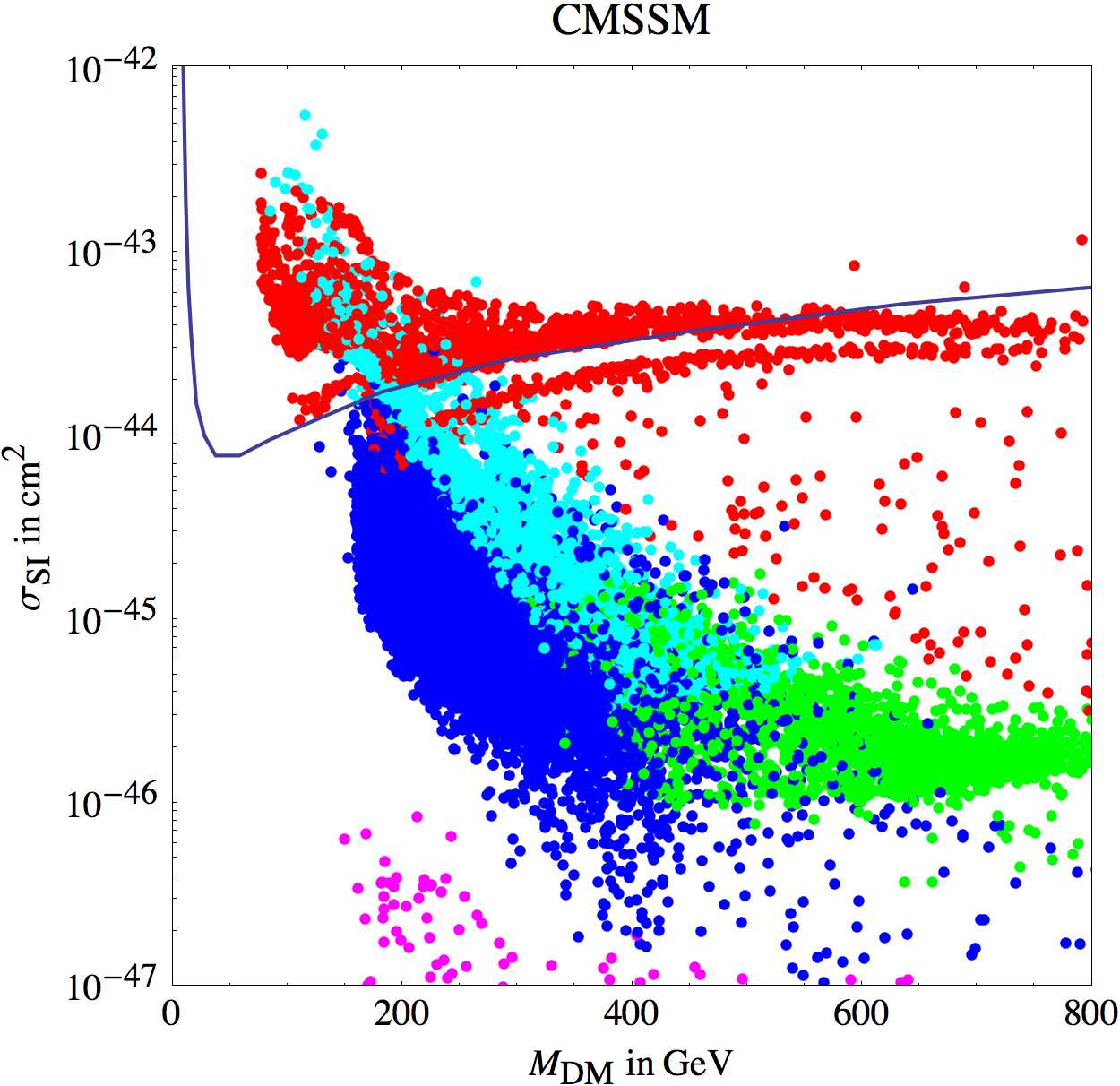}$$\vspace{-1cm}
\caption{Global CMSSM fit updated to 165/pb of ATLAS data.
\label{fig:fit}}
\end{figure}

Fig.\fig{fit}a shows the results: LHC has a minor impact, eliminating
the part of the best-fit parameter space with lighter sparticles.
Furthermore the recent Xenon-100 search for direct detection of Dark Matter~\cite{Xenon10010} disfavors the
region in pink in fig.\fig{fit}a and b.
To understand what is this region now disfavored,
fig.\fig{fit}c shows that the best CMSSM fit  is made of a few qualitatively different corners of the CMSSM parameter space,
corresponding to different ad hoc mechanisms that allow to reproduce the DM abundance:
\begin{itemize}
\item 
The now disfavored red dots  correspond to the
``well tempered bino-higgsino'' mechanism i.e.\ $M_1 \sim |\mu|$,  that in the CMSSM is possible for large $m_0$.

\item The most natural DM annihilation mechanism (neutralino annihilations into sleptons) has been excluded because light enough sleptons are
no longer allowed within the CMSSM.

\item   The ``higgs-resonance'' mechanism for enhanced DM annihilation, i.e.\ light $M_{\rm DM} \approx m_h/2$ and consequently a relatively light gluino,
was allowed in the previous global fit~\cite{fit}.  It has now been excluded by the new ATLAS CMSSM bounds~\cite{SUSYLHC2}.

\item 
The remaining allowed mechanisms are:
slepton co-annihilations (blue dots), $H$ or $A$ resonance (green dots),
$h,H,A$ mediation at large $\tan\beta$ (cyan dots), stop co-annihilations (magenta dots).
\end{itemize}


\section*{References}
\begin{thebibliography}{99}


\bibitem{add}
N.~Arkani-Hamed, S.~Dimopoulos and G.~R.~Dvali,
Phys.\ Lett.\ B {429} (1998) 263.

\bibitem{jj}
ATLAS collaboration, Phys. Lett. B694 (2011) 327 [arXiv:1009.5069].
CMS collaboration, Phys. Rev. Lett. 105 (2010) 262001 [arXiv:1010.0203].

\bibitem{LHCd}
\hepart[1101.4919]{R. Franceschini et al.}.

\bibitem{SUSYLHC}
\hepart[1101.1628]{CMS collaboration}.
\hepart[1102.5290]{ATLAS collaboration}.

\bibitem{SUSYLHC2}
\hepart[1102.2357]{ATLAS collaboration}.
ATLAS collaboration, note ATLAS-CONF-2011-086 (6 june 2011).


\bibitem{BS}
R. Barbieri, A. Strumia, Phys. Lett. B490 (2000) 247 [arXiv:hep-ph/0005203].


\bibitem{LHCnat}
\hepart[1101.2195]{A. Strumia}.

\bibitem{fit} \hepart[1104.3572]{M. Farina et al.}. See references therein for other global CMSSM fits.


\bibitem{Xenon10010}
\hepart[1104.2549]{{\sc Xenon}100 collaboration}.

\end{thebibliography}
\end{document}